\documentclass{aastex63}

\usepackage{amsmath}

\received{}
\revised{}
\accepted{}
\submitjournal{ApJL}

\shorttitle{Superflare light echo from faint debris disk}
\shortauthors{Arimatsu \& Kamizuka}

\begin{document}

\title{Faint debris disk peering through superflare light echo}

\correspondingauthor{Ko Arimatsu}
\email{arimatsu@kwasan.kyoto-u.ac.jp}

\author{Ko Arimatsu}
\affiliation{The Hakubi center/Astronomical Observatory, Graduate School of Science,  Kyoto University \\
Kitashirakawa-oiwake-cho, Sakyo-ku, \\
Kyoto 606-8502, Japan}

\author{Takafumi Kamizuka}
\affiliation{Institute of Astronomy, the University of Tokyo, 2-21-1 Osawa, Mitaka, Tokyo 181-0015, Japan}

\begin{abstract}
We present the detectability of strong mid-infrared (MIR) light echoes from faint debris disks illuminated by bright superflares of M-dwarf stars.
Circumstellar dust grains around an M-dwarf star are simultaneously heated by superflare radiation. 
One can thus expect their re-emission in the MIR wavelength regime.
According to our model calculations for the Proxima Centauri system, the nearest M-dwarf star system, 
thermal emission echos from an inner ($r < 1~{\rm au}$) debris disk with a total mass down to that of the solar system’s zodiacal dust 
are expected to emerge at wavelengths longer than $\sim 10 \, {\rm \mu m}$ with a strength comparable to or greater than a white-light superflare.
Also, observable echos from inner- ($r \lesssim 0.5~{\rm au}$) debris disks irradiated by energetic ($\gtrsim 10^{33.5}~{\rm ergs}$) superflares of nearby ($D < 3~{\rm pc}$) M-dwarfs are expected.
Our simulation results indicate that 
superflare monitoring using high-speed optical instruments like OASES and 
its prompt follow-up using ground-based MIR instruments, such as TAO/MIMIZUKU,
can detect these MIR light echoes from debris disks around solar neighborhood flare stars.
\end{abstract}


\section{Introduction}
Detections and investigations of infrared emissions from debris disks (DDs) provide unique opportunities for probing exoplanetary systems.
These DDs represent planetesimal belts, 
sign-spots of terrestrial planets \citep{Raymond11}. 
Radial distributions of the DD dust thus provide information on planetary system architectures.
Infrared thermal emissions from such DDs around solar neighborhood stars have been investigated through photometric, spectroscopic, and interferometric observations. 
So far, these observations have detected infrared dust emission components among $\sim 20\%$ of nearby main-sequence stars \citep{Wyatt08}.  
The observed components correspond to dust belts with masses much higher than solar systems' zodiacal dust clouds or those with much higher temperatures residing in the innermost side of the planetary systems  \citep{Kral17}. 
Detections of cold or warm DDs with masses comparable to the solar system's zodiacal cloud remain one of the ultimate goals for exoplanetary science.

Among main-sequence stellar populations, M-dwarf stars are the most common in the Galaxy.
Investigating the circumstellar environment around this population would provide common characteristics of planetary systems.
However, the circumstellar dust environment around M-dwarfs is poorly understood, except for some extreme cases (e.g., \cite{Kalas04}), due to the observational difficulty of faint infrared dust emission irradiated by relatively weak stellar radiation.
Proxima Centauri (hereafter Proxima) is the closest ($D = 1.3$ pc; \cite{vanLeeuwen07}) and the most extensively studied M-dwarf (M5.5V) star.
This M-dwarf stellar system likely has two planets.
One is a likely terrestrial planet named Proxima b \citep{Anglada16} rotating in a possible habitable zone (semimajor axis of $\sim 0.05\, {\rm au}$).
Another is a more massive planet in a broader orbit ($\sim 1.5\, {\rm au}$), recently discovered by \citet{Damasso20}. 
Investigating circumstellar dust emission around this well-studied system should provide fundamental knowledge of relationships between planets and DDs in M-dwarf systems.
\citet{Anglada17} reported sub-mm excess emission possibly from circumstellar dust forming three discrete ring-like structures; inner- (radius $r \sim 0.4~{\rm au}$), middle- ($1-4~{\rm au}$), and outer-radius ($30~{\rm au}$) components.
If the emission features are real, they suggest a complex DD system formed and evolved around Proxima.
However, re-analyses of the sum-mm data by \citet{MacGregor18} found that the middle-radius ring component was a false detection due to flux variation of Proxima.
\citet{MacGregor18} also argued about possible false detections of inner and outer components due to unstable stellar activities and contamination of background interstellar dust emission, respectively.

As reviewed above, discoveries and investigations of faint DDs around M-dwarfs are challenging, even for the closest target,
mainly due to low stellar luminosity
and unstable stellar activities.
This {\it Letter} proposes a new observation method of DDs around M-dwarfs peering through the superflare activity.
Some M-dwarfs, including Proxima, are known for their intensive and frequent superflare activities with typical timescales of $\sim$ minutes (e.g., \cite{Howard18}).
Since bright superflares irradiate circumstellar dust grains, thermal re-emission is expected at mid-infrared (MIR) wavelength ranges.
This process has not been examined, possibly due to observational difficulties of a sub-minute-scale transient event at the optical and MIR regimes. 
The recent development of optical high-cadence observation systems (e.g., \cite{Arimatsu17}) enables us to make real-time detections of very short timescale astronomical transients like superflare.
Furthermore, TAO/MIMIZUKU \citep{Kamizuka14}, a ground-based MIR instrument for studying time variability events, will begin its operations in the early 2020s.
This study aims to examine detectability and typical observational characteristics of MIR re-emission from faint DDs irradiated by superflare radiations for the first time,
based on simple model calculations for the Proxima system.

\section{Models}

We assume hypothetical circumstellar dust particles uniformly distributed in a single ring structure without height around Proxima as our fiducial model. 
In this model, we adopt the standard grain size distribution ${\rm d}N \propto a^{-3.5} ~{\rm d}a$ \citep{Dohnanyi69}.
The upper limit of the grain radius is arbitrary set to be $a_{\rm max} = 1000\, {\rm \mu m}$.
On the other hand, we set the lower limit radius $a_{\rm min} = 1\, {\rm \mu m}$, where dust grains are blown out by radiation pressure or stellar wind of M-dwarfs \citep{Augereau06}.
Total dust mass is set to be $1 \times 10^{-8}~M_\oplus$, comparable to the total mass of the solar system's zodiacal dust \citep{Nesvorny10}.

In this study, each dust grain smaller than $a_{\rm max}$ is approximated to be in instantaneous thermal equilibrium with the stellar radiation in a timescale much shorter than a superflare event ($\sim$ minutes).
This assumption is plausible under a realistic range of thermal inertia of dust grains ($\Gamma \ll 10^5 \, {\rm J \, m^{-2} \, s^{-1/2} \, K^{-1}}$).
Equilibrium temperature at $t$ (the time since the beginning of the flare, measured in the observer’s frame), $T_{\rm eq}(t)$, for a spherical dust grain at a radius $a$ is thus determined by the solution of the following equation,

\begin{equation}
\int_0^{+\infty}  4 \pi a^2  Q_{\rm abs}(\lambda,a) \pi B(\lambda,T_{\rm eq}(t)) ~{\rm d}\lambda = 
\int_0^{+\infty} Q_{\rm abs}(\lambda,a)~ \pi a^2 J_*(\lambda,r,\phi,t) ~{\rm d}\lambda, 
\end{equation}
where $Q_{\rm abs}(\lambda,a)$, $B(\lambda,T_{\rm eq}(t))$, and $J_*(\lambda,r,t)$ are 
the absorption efficiency for a dust grain, 
the Planck function at grain temperature $T_{\rm eq}(t)$, and the intensity of the stellar radiation field, respectively.
For $Q_{\rm abs}$, we adopt two models.
One is a simple efficiency model, i.e., 
$Q_{\rm abs}(\lambda, a) = 1$ for $\lambda < \lambda_0$ and  $Q_{\rm abs}(\lambda, a) = \lambda_0/ \lambda$ with $\lambda_0 = 2 \pi a$ for $\lambda > \lambda_0$. 
This model is sometimes adopted as a simple approximation of absorption efficiency computed for carbonaceous dust grains by \citet{Laor93} (e.g., \cite{Lestrade09}). We refer to it as the ``carbon model” in this study.
The other is an astronomical silicate model introduced by \citet{Draine84} and \citet{Laor93}. 
The latter model approximates the strong MIR silicate features observed by several warm DDs \citep{Fujiwara12} and extrasolar zodiacal emission studies \citep{Kral17}.
We refer to this model as the ``silicate model'' in this study. 

\begin{figure}[!pt]
\begin{center}
   \includegraphics[scale=1]{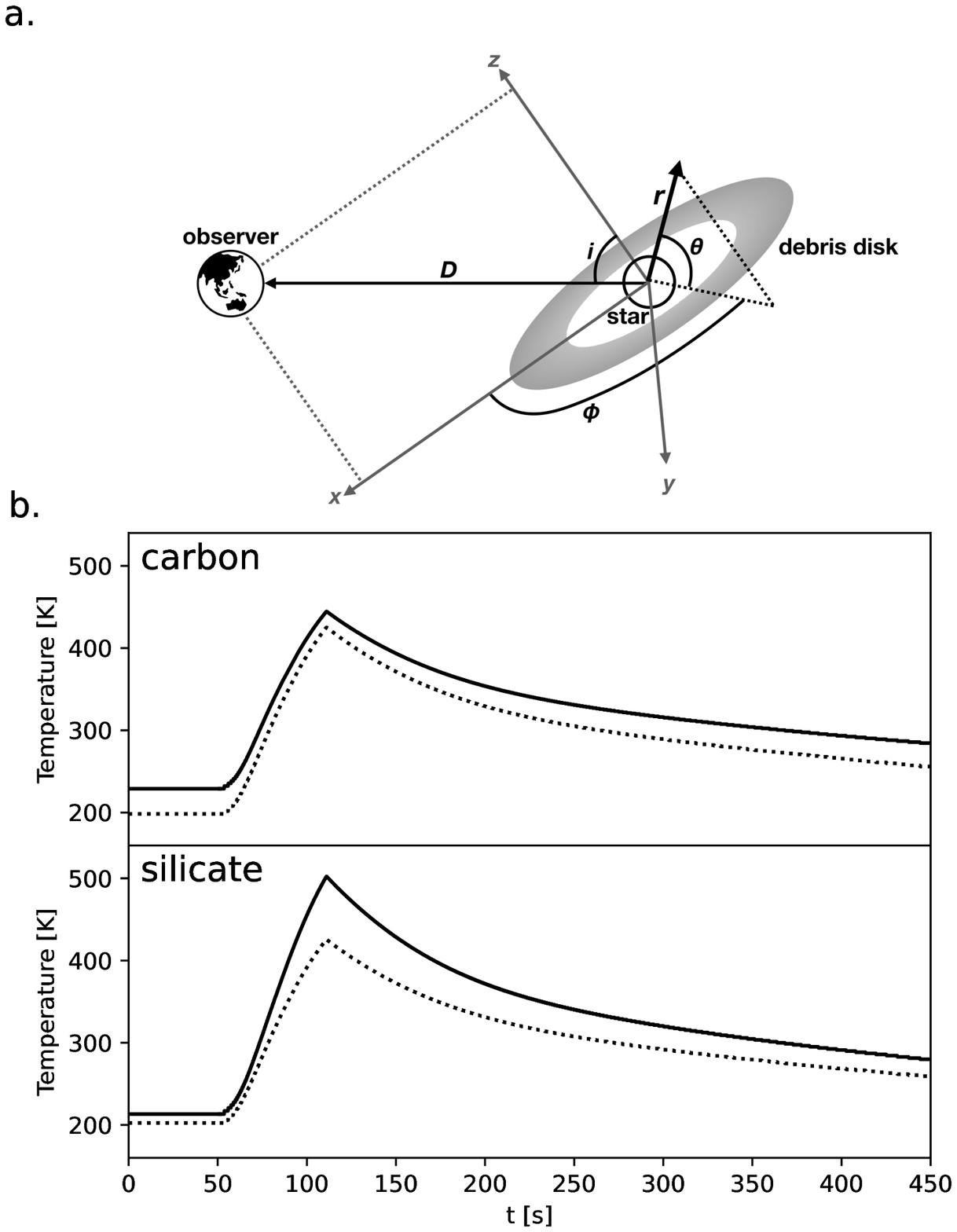}
   \caption{(a) notations used in the present study. (b) simulated temperatures of circumstellar dust grains with $a = 1 ~{\rm \mu m}$ (solid line) and $a = 1000~{\rm \mu m}$ (dashed line) as a function of time $t$. 
   Upper and lower panels show apparent temperature of featureless (carbon) and silicate grains located at $r = 0.35~{\rm au}$, $i = 45\degr$, and $\phi = 0$, respectively.
   $t = 0$ is the beginning time of the superflare event.
  }
 \label{fig0}
\end{center}
\end{figure}

In the following, we define $\phi$, the angle between a direction vector pointing from the stellar center to the direction towards the observer in the ring plane 
(corresponding to $x-y$ plane in Figure~\ref{fig0}a)
and that to the position under consideration.
For simplicity, we assume that the flare appears at $\phi = 0$ in the stellar surface. 
$J_*(\lambda,r,\phi,t)$ in the ring plane is then given by 
\begin{equation}
J_*(\lambda,r,\phi,t) = J_{\rm qui}(\lambda,r) + L(r,\phi ,t') \, B(\lambda,T_{\rm fla}),
\end{equation}
where $J_{\rm qui}(\lambda,r)$ and $L(r,\phi ,t')$ are the intensity of the stellar radiation field during the quiescent state of Proxima and the scale factor for the intensity of the flare.
In this simulation, we assume the spectral energy distribution of the flare to be a single-temperature ($T_{\rm fla}$) blackbody radiation, i.e., $B(\lambda,T_{\rm fla})$.
When we observe the circumstellar structures, we detect those that are affected by the radiation field at different times $t'$ due to the optical path difference given by
\begin{equation}
t' = t - \frac{r}{c} (1 - \cos{\phi} \, \sin{i}),
\end{equation}
where $c$ is the speed of light
and $i$ is the inclination of the ring, i.e., the angle between the rotational axis of the DD ($z$-axis in Figure~\ref{fig0}a) and the direction towards the observer.
For $J_{\rm qui}(\lambda,r)$, we adopt the full spectral energy distribution template of Proxima provided by \citet{Ribas17}.
For $L(r,\phi,t')$, we use a light curve profile template proposed by \citet{Davenport14}.
According to observations of a bright superflare event by \citet{Howard18}, the light curve full-time width at half the maximum flux ($t_{\rm 1/2}$) appears to be shorter than 2 minutes (corresponding to their sampling timescale). 
We thus set $t_{\rm 1/2}$ to be 1 minute.
A white-light flare temperature $T_{\rm fla}$ is set to be  $T_{\rm fla} = 9000 \, {\rm K}$.
Since the white-light flare is expected to appear in a small but finite area of the stellar surface located at $\phi = 0$, 
we assume the following equations
\begin{equation}
L(r,\phi ,t')= 
  \begin{cases}
    L(r,\phi = 0 ,t') \, \cos{\phi}, & -\pi/2 \leq \phi \leq \pi/2\\
    0, & \phi < -\pi/2,  \phi > \pi/2.
  \end{cases}
\end{equation}
This study assumes a white-light superflare with a total bolometric energy $E_{\rm tot}$ of $E_{\rm tot} = 10^{33.5}$ ergs, corresponding to the \citet{Howard18} bright superflare event.
$L(r,\phi ,t)$ is thus normalized to satisfy the following equation:
\begin{equation}
E_{\rm tot}  = \int^{\pi/2}_{-\pi/2} \int^{\pi/2}_{-\pi/2} \int^{\infty}_{-\infty} \int^{\infty}_{0} L(r,\phi = 0 ,t') \, B(\lambda,T_{\rm fla}) \, r^2 \cos^2{\theta} \ \cos{\phi} \ {\rm d}\lambda \, {\rm d}t' \,{\rm d}\theta \, {\rm d}\phi,
\end{equation}
where $\theta$ is the angle with respect to the ring plane.

Figure~\ref{fig0}b shows the simulated temperatures of dust grains ($T_{\rm eq}(t)$) located at $r = 0.35~{\rm au}$, $i = 45\degr$ (corresponding to the inner-radius component by \cite{Anglada17}), and $\phi = 0$ as a function of $t$.
In the quiescent state, $T_{\rm eq}$ for carbon and silicate grains are around $200~{\rm K}$.
For $t > 50~{\rm s}$, which corresponds to $t' > 0~{\rm s}$ for the assumed location, $T_{\rm eq}(t)$ drastically raises as the dust grain irradiated by the superflare.
In both cases, the smaller ($a = 1~{\rm \mu m}$) grains reach higher $T_{\rm eq}(t)$ than the larger ($a = 1000~{\rm \mu m}$) grains due to relatively small $Q_{\rm abs}$ at longer wavelength ranges.
We should note that the peak temperatures are still much lower than the sublimation temperature ($T = 1000-2000~{\rm K}$; e.g., \cite{Sezestre19}) under the assumed conditions.
The total emission intensity of the DD is obtained by integrating the fluxes of the entire volume elements under the assumption that the disk is optically thin.

\section{Results}

\begin{figure}[!pt]
\begin{center}
   \includegraphics[scale=0.85]{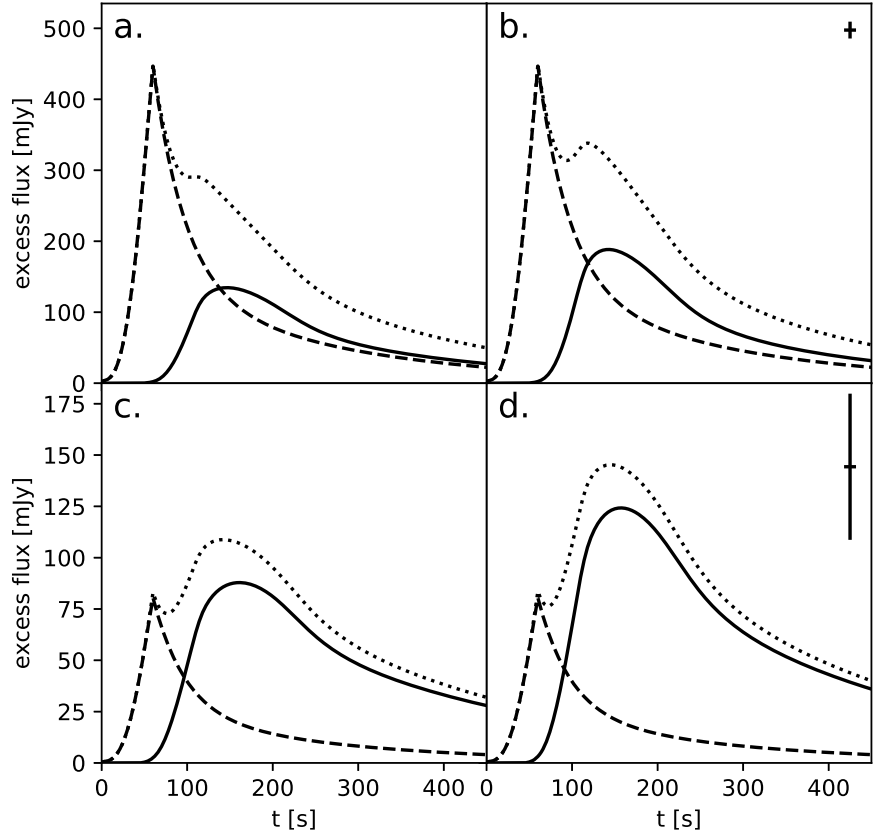}
   \caption{Simulated light curve of the white-light flare (dashed line), the MIR light echo from the DD (solid line), and the total excess (dotted line), respectively.
   Panels (a) and (b) show simulated light curves for the DDs with carbon and silicate grains at the $11.6~{\rm \mu m}$ band, respectively.
   In the quiescent state, the total stellar + DD fluxes are 801 and 800 mJy for carbon and silicate grains, respectively.
   Panels (c) and (d) are the same as (a) and (b), but at the $20.8~{\rm \mu m}$ band,
   where the quiescent-state total fluxes are 304 and 308 mJy for carbon and silicate grains, respectively.
   Expected $1\sigma$ photometric uncertainties for the TAO/MIMIZUKU $11.6~{\rm \mu m}$ and $20.8~{\rm \mu m}$ bands with an integration time of 10~seconds are also shown as crosses in (b) and (d), respectively.
  }
 \label{fig1}
\end{center}
\end{figure}

Figure~\ref{fig1} shows the simulated light curves of the excess emission from the model DD with carbon (panels a and c) and silicate (b and d) dust grains at MIR $11.6~{\rm \mu m}$ and $20.8~{\rm \mu m}$ bands, overlaid with the white-light flare component.
These simulated bands correspond to MIR atmospheric windows and are adopted for photometric bands of ground-based instruments (e.g., \cite{Kamizuka14}).
The quiescent-state stellar photospheric + DD fluxes are $\sim 800$ and
$\sim 300$ mJy at the $11.6~{\rm \mu m}$ and $20.8~{\rm \mu m}$ bands, respectively.
We should note that the photometric errors of objects with the total fluxes into consideration at the MIR bands are dominated by the thermal background noise. 
The contribution of the quiescent-state radiation is not expected to be attributed to the photometric accuracy. 
In this simulation, the inner and outer radii of the DD ring are set to be 0.3 and 0.4~au, respectively, which are consistent with the radius of the inner-radius DD component proposed by the previous sub-mm observations \citep{Anglada17}. 
The inclination is set to be $i = 45\degr$, following \citet{Anglada17}.
The MIR excess from the DD reaches the observer $\sim 60\, {\rm s}$ after the beginning time of the rise phase of the superflare event.
At the $11.6~{\rm \mu m}$ band (Figure~\ref{fig1}a and b), a strong emission peak after the white-light flare appears in the light curve of the total excess emission for both material cases.
The amplitudes of the excess is greater than a reference instrument detection limit (Figure~\ref{fig1}b; \cite{Kamizuka14}).
On the other hand, at the $20.8~{\rm \mu m}$ band (Figure~\ref{fig1}c and d), the amplitudes of the MIR echo peak exceed that of the white-light flare emission but are marginally higher than the instrumental detection limit (Figure~\ref{fig1}d) because of the smaller transparency of Earth's atmosphere. 
In addition, at the $20.8~{\rm \mu m}$ band, a significant flux error of the quiescent-state emission, which needs to be subtracted precisely but marginally detectable, is expected to contribute to the uncertainty of the observed excess fluxes.

\begin{figure}[!pt]
\begin{center}
   \includegraphics[scale=0.85]{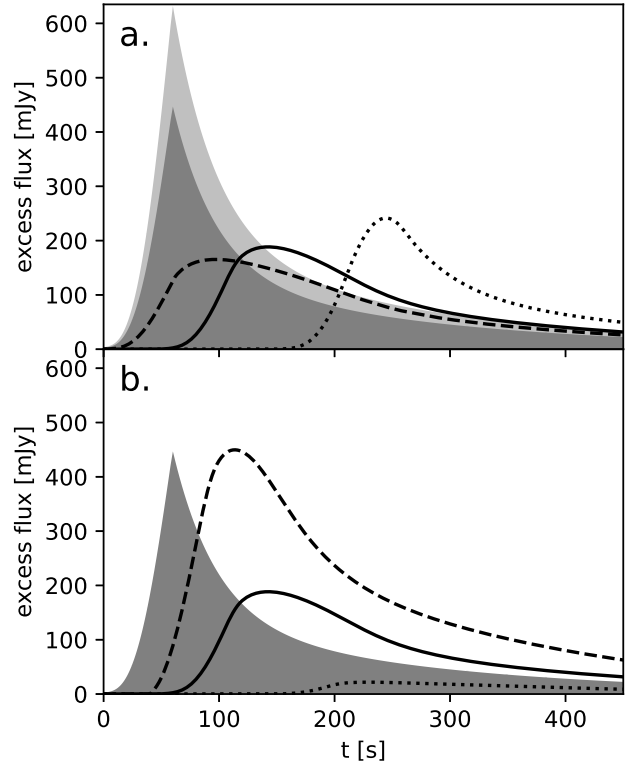}
   \caption{Panel (a): Simulated light curves of the $11.6~{\rm \mu m}$ band echo from the DDs with inclinations of $0\degr$ (face-on, dotted line), $45\degr$ (our fiducial model, solid line), and $i = 90\degr$ (edge-on, dashed line), overlaid with the white-light flare light curves (the dark-gray area for $i = 45\degr$ and the light-gray area for $i = 90\degr$), respectively.
   In the $i = 0\degr$ case, the white light flux becomes zero under our assumed conditions (see text).
   We assume a single ring structure of silicate grains with a radius of $0.3-0.4~{\rm au}$.
   Panel (b): Same as the panel (a), but for different radii, $r = 0.15-0.25~{\rm au}$ (dashed line), $0.3-0.4~{\rm au}$ (our fiducial model, solid line), and  $0.9-1~{\rm au}$ (dotted line), respectively.
   For each light curve, we assume a single ring structure of silicate grains with $i = 45\degr$.
   In both panels, total dust mass is set to be $1 \times 10^{-8}~M_\oplus$.
  }
 \label{fig2}
\end{center}
\end{figure}

By applying different models with different inclinations and radial distances, 
the general detectability of the MIR echo emission can be investigated.
Figure~\ref{fig2}a presents how the light curve changes at different inclinations. 
For the face-on case ($i = 0\degr$), a stronger MIR emission peak appears at $t \sim 250~{\rm s}$ caused by synchronizing a delay time of the emission peak for each DD element at the same $r$.
Furthermore, the apparent fluxes of the white light component decrease with $\cos{i}$ because the flare is assumed to occur in the ring plane (Section~2).
In the lower $i$ cases, the excess emission thus can be easily detected with follow-up MIR observations.
For the edge-on case ($i = 90\degr$), a weaker MIR emission peak appears almost simultaneously as the flare peak.
In addition, the apparent white light component is expected to be brighter than the fiducial model case under our assumed conditions.
It is thus difficult to resolve the peak feature of the MIR emission from the white-light flare profile.
However, a more extended thermal excess emission is expected with an amplitude comparable to the white-light flare component at $t > 100~{\rm s}$.
Thus the MIR emission would be observed as a smooth excess feature during the decay time, even in the edge-on case.

Figure~\ref{fig2}b shows the radial distance dependence of the MIR emission light curve at the $11.6~{\rm \mu m}$ band.
In this figure, we present three cases. 
In each case, a single ring structure with inner and outer radii of 0.15 and 0.25$~{\rm au}$ (inner model), 0.3 and 0.4$~{\rm au}$ (our fiducial model), and 0.9 and 1.0$~{\rm au}$ (outer model) is assumed, respectively. 
For the inner model case, the MIR excess shows a stronger peak than our fiducial model due to the higher temperature ($T_{\rm eq} \sim 600~{\rm K}$) of dust grains.
In this case, the MIR echo emission can be detected as an excess brighter than the white-light flare peak flux.
On the other hand, for the outer model case, the excess intensity becomes fainter than the white-light flare excess flux.
However, the excess level is still comparable to the assumed instrument detection limit under the assumed dust mass ($1 \times 10^{-8}~M_\oplus$).
Thus the MIR light echos from Proxima's DD with its total mass of $\sim 1 \times 10^{-8}~M_\oplus$ would be detectable if it resides within $\sim 1~{\rm au}$ from the central star.

\begin{figure}[!pt]
\begin{center}
   \includegraphics[scale=1]{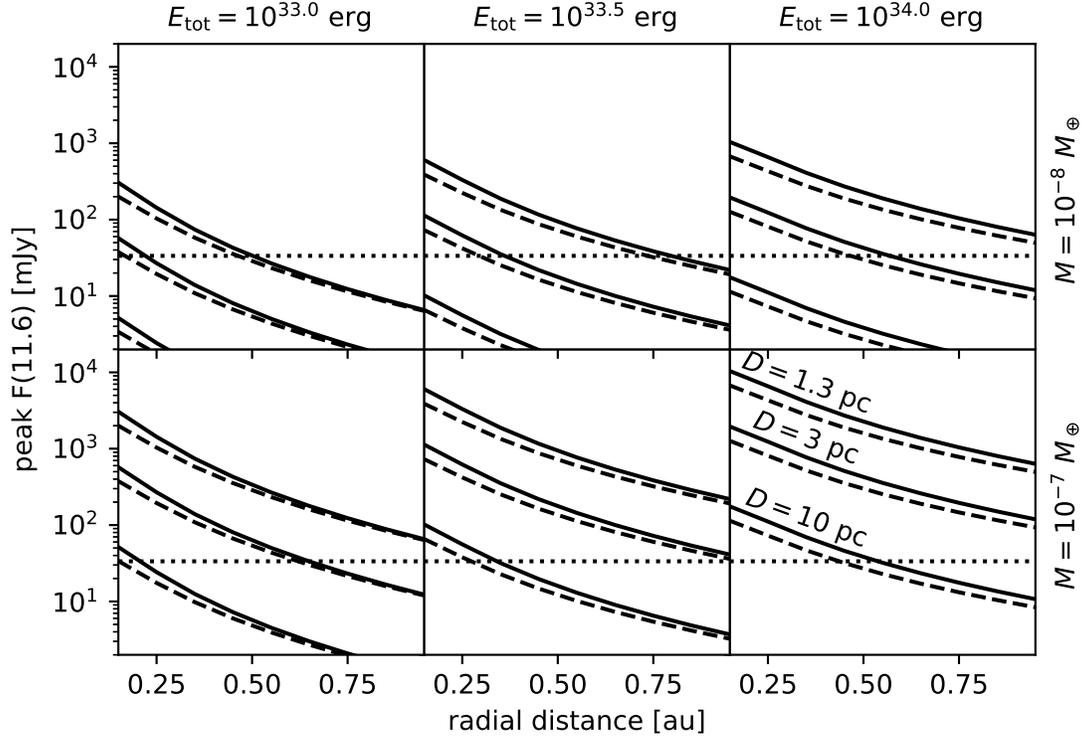}
   \caption{
   Peak fluxes of the $11.6~{\rm \mu m}$ band light echo by a superflare with a total bolometric energy $E_{\rm tot} = 10^{33.0}$ (left), $10^{33.5}$ (middle, our fiducial case), and $10^{34.0}~{\rm ergs}$ (right) as a function of radial distance. The total dust mass of the DD is assumed to be $10^{-8}~M_{\oplus}$ (upper panels, our fiducial case) and $10^{-7}~M_{\oplus}$ (lower panels).
   The Upper, middle, and lower lines in each panel represent the fluxes at $D = 1.3$, 3, and 10~pc, respectively. 
   Dashed and solid lines represent the caerbon and silicate grain models, respectively.
   Horizontal dotted lines represent the $5\sigma$ detection limit for the TAO/MIMIZUKU $11.6~{\rm \mu m}$ band with an integration time of 20~seconds.}
 \label{fig4}
\end{center}
\end{figure}

Figure~\ref{fig4} shows peak fluxes of the $11.6~{\rm \mu m}$ band echos with different radial distances of the disk $r$, total dust masses, flare bolometric energies $E_{\rm tot}$, and distances $D$.
For our fiducial case ($E_{\rm tot} = 10^{33.5}~{\rm ergs}$),
The observable MIR echoes from inner-radii ($\lesssim 0.3\--0.4~{\rm au}$) $10^{-8}~M_\oplus$ DDs around M-dwarfs located within $D \lesssim 3$~pc
are expected.
According to the monitoring studies (e.g., \cite{Gunther20}), the occurrence rate of flares with $E_{\rm tot}$ comparable to or greater than $10^{33.5}~{\rm ergs}$ is approximately $10^{-1.5}~{\rm d^{-1}}$.   
If we assume $E_{\rm tot} = 10^{33.0}~{\rm ergs}$, i.e., less energetic but more frequent (the occurrence rate of $\sim 10^{-1}~{\rm d^{-1}}$) flares, 
the echos from the $D \sim 3$~pc systems become undetectable.
In turn, if we assume more energetic ($E_{\rm tot} = 10^{34.0}~{\rm ergs}$, the corresponding occurrence rate of $\sim 10^{-2}~{\rm d^{-1}}$) flares or more massive ($10^{-8}~M_\oplus$) DDs, observable echoes from more distant (up to $D \sim 10~{\rm pc}$) systems are expected.

\begin{figure}[!pt]
\begin{center}
   \includegraphics[scale=0.85]{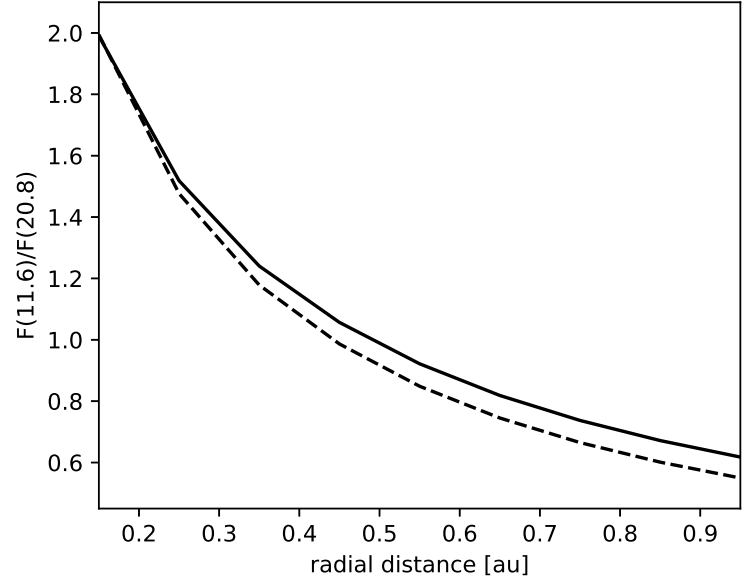}
   \caption{The ratio of the time-averaged fluxes of the excess emission from the DDs at the $11.6~{\rm \mu m}$ and $20.8~{\rm \mu m}$ bands ($F(11.6)/F(20.8)$)  as a function of the radial distance of the DD ring.
 The solid and dashed lines represent $F(11.6)/F(20.8)$ of the DD composed of carbon and silicate grains, respectively.
 The width of the ring annulus is set to be $0.1~{\rm au}$.}
 \label{fig5}
\end{center}
\end{figure}

Assuming the radial distances of the DD to be $< 1~{\rm au}$, the angular size of its echo emission is expected to be smaller than $1\arcsec$ for the most nearby M-dwarfs.
Since these echo emissions would be marginally resolved or unresolved for the current and near-future MIR instruments, 
it is difficult to estimate the geometrical characteristics of the DDs from direct imaging observations.  
On the other hand, as shown in Figures~\ref{fig2}, the delay time of the MIR echo peak is determined by both the inclination angle and the radial distance of the DD. 
We find that multi-band photometry of the MIR light echoes would be useful for estimating the radial distance of the unresolved DDs.
Figure~\ref{fig5} shows ratios of time-averaged excess fluxes at the $11.6~{\rm \mu m}$ and $20.8~{\rm \mu m}$ bands ($F(11.6)/F(20.8)$) against the radial distance of the ring. 
$F(11.6)/F(20.8)$ rapidly decreases with radius as the dust temperature decreases.
Also, $F(11.6)/F(20.8)$ does not highly depend on the grain models taken into consideration in this study.
We should also note that $F(11.6)/F(20.8)$ does not depend on the total dust mass or the inclination of the DD.
By determining the radial distance from the MIR color, one can then estimate the inclination of the DD from the delay time of the MIR echo peak. 
The data from the multi-band high-cadence MIR echo observations would thus provide invaluable new information on the geometrical characteristics of unresolved DDs.

\section{Discussions}

\subsection{Detectability of faint DDs around M-dwarfs through MIR light echoes}
The proposed method requires follow-up MIR observations of second-scale optical transitions on timescales shorter than several minutes. Instruments for optical high-cadence photometry and those dedicated to fast follow-up MIR observations are thus needed. 
Detections of superflare in real-time will be possible with high-cadence optical observation instruments, such as OASES observation systems \citep{Arimatsu17}. 
The OASES systems were initially developed for monitoring stellar occultation events by outer solar system objects \citep{Arimatsu19}. 
However, they were recently used for surveys of short-timescale optical astronomical events with timescales of seconds to sub-seconds \citep{Arimatsu21}. 
With upgrading OASES systems, the beginning of the rising phase of superflares can be detected almost simultaneously. 
Furthermore, we expect that near-future ground-based infrared instruments will achieve fast follow-up detections of MIR light echos from the DDs.
For example, MIMIZUKU \citep{Kamizuka14} mounted on the TAO 6.5-m telescope can potentially start observations about two minutes after detecting a superflare with OASES, taking advantage of the telescope's fast-pointing capability.
MIMIZUKU will enable to carrying out MIR multi-band follow-up observations of the short-timescale echos.
Coordinated observations with OASES and TAO will be useful for exploring faint DDs around nearby flare stars' systems, such as Proxima.

\subsection{Prospects for near future observations of faint DDs}
As reviewed in Section~1, sub-mm DDs observations suffered contaminations with unstable stellar activities and the background interstellar dust emission \citep{MacGregor18}.
The proposed observation method provides a unique opportunity of exploring faint DDs around nearby M-dwarfs free from these contaminations.
Recent interferometric observations revealed circumstellar faint and warm dust around nearby main-sequence stars \citep{Kral17,Ertel20}.
However, targets for these observations are limited to early-type (earlier than K-type) bright stars due to limitations of sensitivities.
Observations of MIR light echoes provide a unique detection method for faint DDs 
around M-dwarfs and can be a complementary study for faint DD surveys with interferometric observations. 

\section{Coclusions}
Our simple model calculations expect bright MIR light echoes from faint DDs around nearby M-dwarfs illuminated by their bright superflares.
Especially for the Proxima system, MIR echos from the inner ($r < 1~{\rm au}$) DDs with total masses down to that of the solar system’s zodiacal dust are expected to be detectable with near-future ground-based instruments.
Fast detections of superflares with optical high-cadence observation systems and their simultaneous follow-up observations with ground-based MIR instruments 
will enable us to detect the MIR echos.
By obtaining the MIR color and the delay time of the echo through the follow-up MIR multi-band observations, the radial distance and inclination angle of unresolved DDs can be obtained.

The present study only focuses on MIR re-emission from circumstellar dust by superflare radiation.
However, optical and ultraviolet light-echo radiation by the reflection of superflare emission is also expected.
The detectability of such re-emission is beyond the scope of this study.
 We plan to test this idea in the future.

\clearpage

\clearpage

\clearpage

\clearpage

\clearpage

\clearpage

\acknowledgments
We thank the anonymous referee for a careful review and for providing constructive suggestions.
The concept of this research was started during the preparation of {\it Monday seminar}, one of the research activities in the department of astronomy, Kyoto University.
We thus appreciate all of the participants of the seminar.
We also thank for Kosuke Namekata for input on superflare researches.
This research has been partly supported by JSPS grants (16K17796, 18K13606, 21H01153).


\begin{thebibliography}{}

\bibitem[Anglada et al.(2017)]{Anglada17} Anglada, G., Amado, P.~J., Ortiz, J.~L., et al.\ 2017, \apjl, 850, L6. doi:10.3847/2041-8213/aa978b

\bibitem[Anglada-Escud{\'e} et al.(2016)]{Anglada16} Anglada-Escud{\'e}, G., Amado, P.~J., Barnes, J., et al.\ 2016, \nat, 536, 437. doi:10.1038/nature19106

\bibitem[Arimatsu et al.(2017)]{Arimatsu17} Arimatsu,~K., Tsumura,~K., Ichikawa,~K., et al.\ 2017, \pasj, 69, 68. doi:10.1093/pasj/psx048

\bibitem[Arimatsu et al.(2019)]{Arimatsu19} Arimatsu,~K., Tsumura,~K., Usui,~F., et al. 2019, Nature Astronomy, 3, 301. doi:10.1038/s41550-018-0685-8 

\bibitem[Arimatsu et al.(2021)]{Arimatsu21} Arimatsu, K., Tsumura, K., Usui, F., et al.\ 2021, \aj, 161, 135. doi:10.3847/1538-3881/abd94d

\bibitem[Augereau \& Beust(2006)]{Augereau06} Augereau, J.-C. \& Beust, H.\ 2006, \aap, 455, 987. doi:10.1051/0004-6361:20054250


\bibitem[Davenport et al.(2014)]{Davenport14} Davenport, J.~R.~A., Hawley, S.~L., Hebb, L., et al.\ 2014, \apj, 797, 122. doi:10.1088/0004-637X/797/2/122

\bibitem[Damasso et al.(2020)]{Damasso20} Damasso, M., Del Sordo, F., Anglada-Escud{\'e}, G., et al.\ 2020, Science Advances, 6, eaax7467. doi:10.1126/sciadv.aax7467

\bibitem[Dohnanyi(1969)]{Dohnanyi69} Dohnanyi, J.~S.\ 1969, \jgr, 74, 2531. doi:10.1029/JB074i010p02531

\bibitem[Draine \& Lee(1984)]{Draine84} Draine, B.~T. \& Lee, H.~M.\ 1984, \apj, 285, 89. doi:10.1086/162480

\bibitem[Ertel et al.(2020)]{Ertel20} Ertel, S., Defr{\`e}re, D., Hinz, P., et al.\ 2020, \aj, 159, 177. doi:10.3847/1538-3881/ab7817

\bibitem[Fujiwara et al.(2012)]{Fujiwara12} Fujiwara, H., Onaka, T., Yamashita, T., et al.\ 2012, \apjl, 749, L29. doi:10.1088/2041-8205/749/2/L29

\bibitem[G{\"u}nther et al.(2020)]{Gunther20} G{\"u}nther, M.~N., Zhan, Z., Seager, S., et al.\ 2020, \aj, 159, 60. doi:10.3847/1538-3881/ab5d3a

\bibitem[Howard et al.(2018)]{Howard18} Howard, W.~S., Tilley, M.~A., Corbett, H., et al.\ 2018, \apjl, 860, L30. doi:10.3847/2041-8213/aacaf3

\bibitem[Kalas et al.(2004)]{Kalas04} Kalas, P., Liu, M.~C., \& Matthews, B.~C.\ 2004, Science, 303, 1990. doi:10.1126/science.1093420

\bibitem[Kamizuka et al.(2014)]{Kamizuka14} Kamizuka, T., Miyata, T., Sako, S., et al.\ 2014, \procspie, 9147, 91473C. doi:10.1117/12.2056184

\bibitem[Kral et al.(2017)]{Kral17} Kral, Q., Krivov, A.~V., Defr{\`e}re, D., et al.\ 2017, The Astronomical Review, 13, 69. doi:10.1080/21672857.2017.1353202

\bibitem[Laor \& Draine(1993)]{Laor93} Laor, A. \& Draine, B.~T.\ 1993, \apj, 402, 441. doi:10.1086/172149

\bibitem[Lestrade et al.(2009)]{Lestrade09} Lestrade, J.-F., Wyatt, M.~C., Bertoldi, F., et al.\ 2009, \aap, 506, 1455. doi:10.1051/0004-6361/200912306

\bibitem[MacGregor et al.(2018)]{MacGregor18} MacGregor, M.~A., Weinberger, A.~J., Wilner, D.~J., et al.\ 2018, \apjl, 855, L2. doi:10.3847/2041-8213/aaad6b

\bibitem[Nesvorn{\'y} et al.(2010)]{Nesvorny10} Nesvorn{\'y}, D., Jenniskens, P., Levison, H.~F., et al.\ 2010, \apj, 713, 816. doi:10.1088/0004-637X/713/2/816

\bibitem[Raymond et al.(2011)]{Raymond11} Raymond, S.~N., Armitage, P.~J., Moro-Mart{\'\i}n, A., et al.\ 2011, \aap, 530, A62. doi:10.1051/0004-6361/201116456

\bibitem[Ribas et al.(2017)]{Ribas17} Ribas, I., Gregg, M.~D., Boyajian, T.~S., et al.\ 2017, \aap, 603, A58. doi:10.1051/0004-6361/201730582


\bibitem[Sezestre et al.(2019)]{Sezestre19} Sezestre, {\'E}., Augereau, J.-C., \& Th{\'e}bault, P.\ 2019, \aap, 626, A2. doi:10.1051/0004-6361/201935250


\bibitem[Stelzer et al.(2013)]{Stelzer13} Stelzer, B., Marino, A., Micela, G., et al.\ 2013, \mnras, 431, 2063. doi:10.1093/mnras/stt225


\bibitem[van Leeuwen(2007)]{vanLeeuwen07} van Leeuwen, F.\ 2007, \aap, 474, 653. doi:10.1051/0004-6361:20078357

\bibitem[Wyatt(2008)]{Wyatt08} Wyatt, M.~C.\ 2008, \araa, 46, 339. doi:10.1146/annurev.astro.45.051806.110525

\end{thebibliography}
\bibliographystyle{aasjournal}

\end{document}